# A digital interface for Gaussian relay networks: lifting codes from the discrete superposition model to Gaussian relay networks

M. Anand and P. R. Kumar
Dept. of ECE and CSL, Univ. of Illinois, Urbana, IL 61801, USA
Email: {amurali2, prkumar}@illinois.edu

*Abstract*—For every Gaussian relay network with a single source-destination pair, it is known that there exists a corresponding deterministic network called the discrete superposition network [1] that approximates its capacity uniformly over all SNR's to within a bounded number of bits. The next step in this program of rigorous approximation is to determine whether coding schemes for discrete superposition models can be lifted to Gaussian relay networks with a bounded rate loss independent of SNR. We establish precisely this property and show that the superposition model can thus serve as a strong surrogate for designing codes for Gaussian relay networks.

We show that a code for a Gaussian relay network, with a single source-destination pair and multiple relay nodes, can be designed from any code for the corresponding discrete superposition network simply by pruning it. In comparison to the rate of the discrete superposition network's code, the rate of the Gaussian network's code only reduces at most by a constant that is a function only of the number of nodes in the network and independent of channel gains.

This result is also applicable for coding schemes for MIMO Gaussian relay networks, with the reduction depending additionally on the number of antennas.

Hence, the discrete superposition model can serve as a digital interface for operating Gaussian relay networks.

## I. Introduction

Computing the capacity of a general Gaussian network is a formidable problem. This has motivated the pioneering work of [3] which aims at developing a class of deterministic networks that can serve as surrogates for Gaussian networks.

The first issue that needs to be settled in such a program is to show rigorously that deterministic networks can indeed approximate the capacity of Gaussian relay networks. This was done in [2] and [1] via different models. The model used in [1], called a discrete superposition model, is a more discrete model in the sense that channel gains and inputs are discrete valued, while they are complex valued in [2].

The next issue in this program of investigation is to rigorously establish a correspondence between coding schemes for Gaussian and superposition networks. Here we show that coding strategies for the discrete superposition model can also be lifted to the Gaussian relay network in such a way that they continue to provide comparable performance. The lifting procedure is particularly simple and consists essentially of just pruning the codewords and using jointly typical decoding.

The discrete superposition model thus provides a *digital interface* for operating the Gaussian network since any coding scheme for the superposition model defines a finite set of transmit signals at every node in the Gaussian network, a finite set of signals for decoding its received signal to, and a mapping from decoded signals to transmit signals. Operating the Gaussian network on this precise digital interface will achieve rates that are close to its capacity if the original scheme for the discrete network does so, thereby providing a framework for designing optimal strategies.

The superposition model may potentially be easier to design codes for than the Gaussian model since noise has been eliminated from the network and the set of inputs and outputs are finite. Potentially, perhaps, wireless network coding could be useful too in studying superposition networks, and subsequently Gaussian networks.

### A. Summary of previous work

The linear deterministic model, introduced in [3], captured broadcast and interference in a wireless network, and the effect of noise in a Gaussian network. Networks constructed with this model approximate the capacity of the original Gaussian network for certain examples like multiple-access [3], broadcast [3], and the two-user interference channels [4].

However, the linear deterministic model cannot capture the phase of a channel gain [1] and also substantially reduces the received signal power at a node [1], [2], due to either of which relay networks constructed with this model can have arbitrarily lower capacity than the original Gaussian network. Hence, the linear deterministic model fails to approximate the capacity of Gaussian relay networks.

## II. Model

### A. System model

We begin by describing the class of Gaussian relay networks of interest. We consider a wireless network represented as a directed graph $(\mathcal{V}, \mathcal{E})$, where $\mathcal{V} = \{0, 1, \ldots, M\}$ represents the set of nodes, and the directed edges in $\mathcal{E}$ correspond to wireless links. Denote by $h_{ij}$ the complex channel gain for link $(i,j) \in \mathcal{E}$. Let the complex number $x_i$ denote the transmission

This material is based upon work partially supported by NSF under Contract No. CNS-0905397, USARO under Contract Nos. W911NF-08-1-0238 and W-911-NF-0710287, and AFOSR under contract No. FA9550-09-0121.

of node $i$. Every node has an average power constraint, taken to be 1. Node $j$ receives

$$y_j = \sum_{i \in \mathcal{N}(j)} h_{ij} x_i + z_j, \quad (1)$$

where $\mathcal{N}(j)$ is the set of its neighbors, $z_j$ is complex white Gaussian noise, $\mathcal{CN}(0,1)$, independent of transmitted signals, and $h_{ij} = h_{ijR} + \imath h_{ijI}$.

### B. Discrete superposition model

We associate with the above Gaussian relay network a deterministic model, as follows. Let

$$n := \max_{(i,j) \in \mathcal{E}} \max\{\lfloor \log |h_{ijR}| \rfloor, \lfloor \log |h_{ijI}| \rfloor\}. \quad (2)$$

The inputs are complex valued, and both real and imaginary parts take values from the $2^n$ equally spaced discrete points $\{0, 2^{-n}, \ldots, 1 - 2^{-n}\}$. The real or imaginary part of an input can be represented in terms of its binary representation $x = \sum_{k=1}^{n} 2^{-k} x(k)$, with each $x(i) \in \mathbb{F}_2$.

The real and imaginary parts of channel gains in the Gaussian network are *quantized* to integers by neglecting their fractional parts. $h'_{ij}$ below denotes the quantized channel gain for link $(i, j)$:

$$h'_{ij} := [h_{ij}] := \text{sign}(h_{ijR})\lfloor |h_{ijR}| \rfloor + \imath\, \text{sign}(h_{ijI})\lfloor |h_{ijI}| \rfloor. \quad (3)$$

The channel between two nodes in the discrete superposition network multiplies the input by the corresponding channel gain and quantizes the product by neglecting the fractional components of both real and imaginary parts, i.e., it forms $[h'_{ij} x_i]$. The outputs of all incoming channels at a receiver node lie in $\mathbb{Z} + \imath \mathbb{Z}$. All the quantized outputs are added up at a receiver by the standard summation over $\mathbb{Z} + \imath \mathbb{Z}$. The received signal at node $j$ is given by

$$y'_j = \sum_{i \in \mathcal{N}(j)} [h'_{ij} x_i]. \quad (4)$$

This model retains the essential superposition property of the wireless channel. Quantization of channel coefficients does not substantially change the channel matrix in the high SNR limit. Also, the effect of noise is captured by constraining the inputs to positive fractions that can be represented by finite bits and by quantization of the channel output.

An important property of the discrete superposition model is that transmit signals in it satisfy a unit peak power constraint, and are thus also valid for transmission in the Gaussian network. That is, encoder outputs in the discrete superposition network can also be used in the Gaussian network.

A Gaussian relay network is shown in Fig. 1(a), where each wireless link is labeled with its channel gain and we have explicitly indicated the addition of Gaussian noise at every receiver. In comparison, its discrete superposition counterpart in Fig. 1(b) preserves multiplication by channel gains and the superposition property of the channel, with the Gaussian noise replaced by quantization of outputs.

The discrete superposition model was first used in [4] in a sequence of networks that reduced the Gaussian interference channel to a linear deterministic interference channel. It was generalized and given its name in [1].

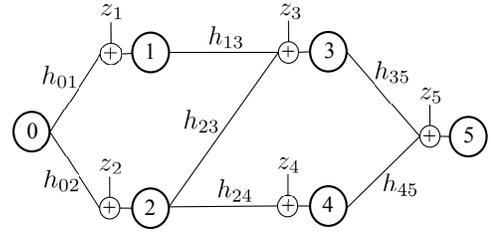

(a) A Gaussian relay network.

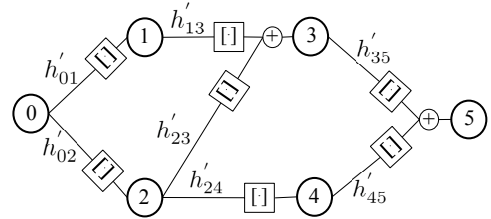

(b) Discrete superposition model of above network.

Fig. 1. Gaussian and discrete superposition models of a network.

## III. LIFTING A CODING SCHEME FOR THE DISCRETE SUPERPOSITION NETWORK TO THE GAUSSIAN NETWORK

A coding strategy for either the Gaussian or superposition relay network specifies codewords transmitted by the source, a mapping from the received signal to a transmit signal for every relay node, and a decoding function for the destination.

We describe how to lift a coding strategy for the discrete superposition network to a strategy for the Gaussian network.

Consider a $(2^{NR}, N)$ code for the discrete superposition network with zero probability of error, for a certain $N$. The probability of error can be reduced to zero due to the deterministic nature of the network; see Sec. IV-A1.

Denote the block of $N$ transmissions at node $j$ in the discrete superposition network by an $N$-dimensional transmit vector $\underline{x}_j$ and similarly the received vector by $\underline{y}'_j$. All signals in the discrete superposition network are a (deterministic) function of the codeword $\underline{x}_0$ transmitted by the source.

Next, we build a $(2^{nNR}, nN)$ code, denoted by $\mathcal{C}_0$, for the discrete superposition network, with every $nN$-length codeword constructed by adjoining $n$ codewords from the old code, for a large $n$. This is again a rate $R$ code since it simply uses the old code $n$ times on the superposition network.

In the $(2^{nNR}, nN)$ code, node $j$
1) breaks up its received signal, denoted by $\underline{y}'_j$, into $n$ blocks of length $N$,
2) applies the mapping used in the $(2^{NR}, N)$ code on each of the $n$ blocks to generate $n$ blocks of transmit signals,
3) and adjoins $n$ blocks of transmit signals to construct a new transmit signal, denoted by $\underline{x}_j$, of length $nN$.

As shown in Fig. 2, the relationship between various signals associated with the transmission of node $j$ is akin to packetization in computer networks.

A subset of the codewords of $\mathcal{C}_0$, defined below, forms the set of codewords of the code for the Gaussian relay network.

Pruning the set of codewords: Node $j$ has a finite set of $\epsilon$-strongly typical $\underline{y}'_j$'s (see [5]) in the code for the superposition network. We randomly, i.e., independently and uniformly, pick a $2^{-n(N\kappa+2\eta)}$ fraction of them and denote the resulting set by $\mathcal{S}_j$. $\kappa > 0$ is defined later in (12) as a function only of the number of nodes in the network and not the channel gains, while $\eta > 0$ is specified later and can be made arbitrarily small. We repeat this pruning procedure for all the nodes.

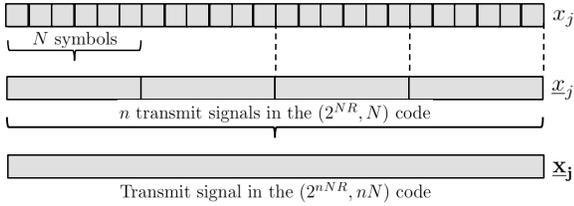

Fig. 2. Relationship among the signals transmitted by node $j$.

Denote the intersection of the inverse images of $\mathcal{S}_j$ in $\mathcal{C}_0$, for $j = 1, 2, \cdots, M$, by $\mathcal{C}_G$. Transmission of any vector in $\mathcal{C}_G$ results in the received vector at node $j$ belonging to $\mathcal{S}_j$ in the discrete superposition network. $\mathcal{C}_G$ forms the set of codewords for the Gaussian network.

Encoding and decoding procedure in the Gaussian network: The source in the Gaussian network transmits a codeword $\mathbf{x_0}$ from $\mathcal{C}_G$. Assume throughout that node 1 can listen, i.e., has a link, to the source. Node 1 receives a noisy signal and decodes to a vector in $\mathcal{S}_1$. We will specify in the sequel how this decoding is to be done. Then, using the encoding function from $\mathcal{C}_0$, it constructs its transmit signal. All nodes operate in a similar way. Finally, the destination decodes its noisy reception to a signal in $\mathcal{S}_M$, and maps it to a codeword by simply using the decoding function from $\mathcal{C}_0$.

Note that we are operating the Gaussian network over the digital interface naturally defined by the signals transmitted and received in the corresponding discrete superposition network.

We summarize the main result concerning the lifting procedure in Theorem 3.1.

*Theorem 3.1:* Consider a Gaussian network with a single source-destination pair and $M-1$ relay nodes, and consider a code for the discrete superposition model of the network that communicates at a rate $R$.

Then, the lifting procedure and the digital interface defined by the discrete superposition model yield a code for the original Gaussian network that communicates at a rate $R - M\kappa$.

It should be noted that $\kappa$ (and also $M$) do not depend on the channel gains. Therefore, the above theorem provides a lifting procedure that attains a rate in the Gaussian network within a bounded amount of $R$ at any SNR.

The theorem applies to any coding scheme for the superposition network and, in particular, to an optimal scheme. Since the capacities of the Gaussian and the superposition network are within a bounded gap [1], the optimal scheme for the superposition network can be lifted to obtain a near-optimal coding scheme for the Gaussian network.

## IV. Proof of Theorem 3.1

Before delving into the details of the proof, we start with a genie-based argument explaining the ideas behind the proof of Theorem 3.1.

Consider the networks in Fig. 1(a) and Fig. 1(b). For simplicity, assume node 1 transmits a symbol $x_1$ and node 2 transmits a symbol $x_2$ (instead of a block of symbols each) from the alphabet for the discrete superposition network. Node 3 receives $y'_3 = [h'_{13}x_1] + [h'_{23}x_2]$ in the discrete superposition network in Fig. 1(b), and it receives $y_3 = h_{13}x_1 + h_{23}x_2 + z_3$ in the Gaussian network in Fig. 1(a). Rewriting $y_3$, we get

$$\begin{aligned} y_3 &= y'_3 + (h_{13}x_1 - h'_{13}x_1) + (h'_{13}x_1 - \lfloor h'_{13}x_1 \rfloor) \quad (5) \\ &\quad + (h_{23}x_1 - h'_{23}x_1) + (h'_{23}x_1 - \lfloor h'_{23}x_1 \rfloor) + z_3 \\ &=: y'_3 + v_3 + z_3. \quad (6) \end{aligned}$$

By definition, $y'_3$ lies in $\mathbb{Z} + \imath\mathbb{Z}$ and is given by

$$y'_3 = [y_3] - [v_3] - [z_3] - c_3, \quad (7)$$

where $c_3$ is the integer part of the carry obtained by adding fractional parts of $v_3$ and $z_3$, i.e., $c_3 := [(v_3 - [v_3]) + (z_3 - [z_3])]$. This gives an upper bound

$$H(Y'_3|Y_3) \leq H([V_3]) + H([Z_3]) + H(C_3). \quad (8)$$

The real and imaginary parts of $[V_3]$ lie in the set $\{-5, -2, \ldots, 2, 5\}$. Hence $H([V_3]) \leq 5$. Similarly, the real and imaginary parts of $[C_3]$ lie in the set $\{-1, 0, 1\}$, and hence $H(C_3) \leq 3$. Since $z_3 = z_{3R} + \imath z_{3I}$ with the real and imaginary parts distributed as independent $\mathcal{N}(0, 1/2)$,

$$\begin{aligned} H([Z_3]) &= 2H([Z_{3R}]) = 2H(Z'_{3R}) \\ &= -2 \sum_{k \in \mathbb{Z}} p_{Z'_{3R}}(k) \log p_{Z'_{3R}}(k) \\ &\leq -2 \sum_{k \in \mathbb{Z}} p_{Z'_{3R}}(k) \log \left( \frac{1}{\sqrt{\pi}} \exp(-(|k|+1)^2) \right) \\ &\leq \log \pi + 2\mathbb{E}(|Z_{3R}|+1)^2 \\ &= \log \pi + 2\mathbb{E}(Z_{3R}^2) + 4\mathbb{E}(|Z_{3R}|) + 2 \\ &\leq \log \pi + 1 + 2\sqrt{2} + 2 < 8. \quad (9) \end{aligned}$$

So $H(Y'_3|Y_3) \leq 16$, and these computations can be repeated for all the nodes in Gaussian network. In general, if there are $M$ incoming signals, rather than just two as in the above example, then

$$\begin{aligned} H(Y'_j|Y_j) &\leq H([V_j]) + H([Z_j]) + H([C_j]) \quad (10) \\ &\leq \log(12M - 2) + 11, \quad (11) \end{aligned}$$

where $v_j$ and $c_j$ are defined, as in (6), with respect to the signal received by node $j$. Let

$$\kappa := \log(12M - 2) + 11 \quad (12)$$

be a function of the total number of nodes and independent of channel gains (or SNR). Now we use a code designed for the superposition network in the Gaussian network. If there were a genie providing $H(Y'_j|Y_j)$ bits of data corresponding to the received signal to node $j$ in every channel use, then node $j$ can recover $\mathbf{y'_j}$ from $\mathbf{y_j}$. Since the genie provides at most $\kappa$ bits to every node, it provides a total of at most $M\kappa$ bits.

Hence, with the genie's aid, a code designed for the discrete superposition network can be used in the Gaussian network at any SNR. Our proof below prunes a fraction of the codewords representing the information that the genie would have provided, so that the decoding can work even without the genie.

## A. The proof

*1) Zero probability of error:* Consider the $(2^{NR}, N)$ code for the superposition network and assume that it has an average probability of error $\delta$, where $0 \leq \delta < 1/2$. Since the superposition network is a noiseless network, each codeword is either always decoded correctly or always decoded incorrectly. Since $\delta < 1/2$, less than half of the codewords are always decoded incorrectly. Discarding them results in a code where all codewords can be successfully decoded, with a small loss in the rate. So, without loss of generality, we assume that the $(2^{NR}, N)$ code (and thus also the $(2^{nNR}, nN)$ code) for the superposition network has zero probability of error.

$\underline{X}_0$, the random variable corresponding to the codeword, has a uniform distribution with $H(\underline{X}_0) = NR$, and induces a distribution on the remaining variables in the network.

*2) Operating over blocks of length $nN$:* In the $(2^{nNR}, nN)$ code, we assume that every node buffers $nN$ of its received symbols, eventually constructing a transmit signal of length $nN$, and transmits it over the next $nN$ channel uses.

For the network in Fig. 1(a), this is possible since nodes can be grouped into levels such that only nodes at one level communicate with another level. For example, nodes 1 and 2 in Fig. 1(a) can buffer their reception till node 0 completes its transmission, then construct their transmit signals, and transmit to nodes 3 and 4 over the next $nN$ channel uses.

For a general network, we need to differentiate between signals received by a node at various time instants to account for causality in construction of their transmit signals. This requires slightly modifying the procedure; see Sec. IV-A6.

*3) Pruning the code with respect to node 1:* Each $\underline{\mathbf{y}}'_j$ (or $\underline{\mathbf{x}}_j$) in $\mathcal{C}_0$ is generated by $n$ independent samples from the distribution of $\underline{Y}'_j$ (or $\underline{X}_j$). Choose $\epsilon > 0$. For a sufficiently large $n$, node 1 has a collection of at most $2^{n(H(\underline{Y}'_1)+\epsilon_2)}$ and at least $2^{n(H(\underline{Y}'_1)-\epsilon_2)}$ $\epsilon$-strongly typical received vectors in the discrete superposition network corresponding to $\mathcal{C}_0$ (see [5]), where $\epsilon_2 > 0$. As $\epsilon \to 0$, $\epsilon_2 \to 0$. With $\eta$ set to $\epsilon_2$, we construct $\mathcal{S}_1$ by randomly selecting a $2^{-n(N\kappa+2\eta)}$ fraction of this collection. We do this by choosing a subset uniformly among all the subsets of the appropriate size. $|\mathcal{S}_1|$ can be upper bounded as follows (see (10)–(12)):

$$\begin{aligned}|\mathcal{S}_1| &\leq 2^{n(H(\underline{Y}'_1)+\epsilon_2)} 2^{-n(N\kappa+2\eta)} \\ &\leq 2^{n(H(\underline{Y}'_1)-H(\underline{Y}'_1|\underline{Y}_1)-\epsilon_2)} = 2^{n(I(\underline{Y}'_1;\underline{Y}_1)-\epsilon_2)}.\end{aligned}$$

Similarly, we can show that $|\mathcal{S}_1| \geq 2^{n(H(\underline{Y}'_1)-N\kappa-3\epsilon_2)}$.

For a large $n$, the number of codewords in $\mathcal{C}_0$ jointly $\epsilon$-strongly typical with a particular $\underline{\mathbf{y}}'_1$ can be bounded independently of the chosen $\underline{\mathbf{y}}'_1$; see [5]. The desired set has $2^{n(H(\underline{X}_0|\underline{Y}'_1)\pm\epsilon_2)}$ codewords for a particular $\underline{\mathbf{y}}'_1$, i.e., transmission of one of those codewords in the superposition network results in node 1 receiving the chosen $\underline{\mathbf{y}}'_1$. Due to the deterministic nature of the channel, the sets of codewords in $\mathcal{C}_0$ jointly typical with two different vectors in $\mathcal{S}_1$ form disjoint sets. To construct $\mathcal{C}_{0,1}$, we pick the set of all codewords in $\mathcal{C}_0$ that are jointly $\epsilon$-strongly typical with some vector in $\mathcal{S}_1$. We have,

$$\begin{aligned}|\mathcal{C}_{0,1}| &= \sum_{\underline{\mathbf{y}}'_1 \in \mathcal{S}_1} \begin{pmatrix}\text{\# of codewords in } \mathcal{C}_0 \text{ jointly}\\ \epsilon\text{-strongly typical with } \underline{\mathbf{y}}'_1\end{pmatrix} \\ &\leq \sum_{\underline{\mathbf{y}}'_1 \in \mathcal{S}_1} 2^{n(H(\underline{X}_0|\underline{Y}'_1)+\epsilon_2)} \quad (13)\\ &\leq 2^{n(H(\underline{Y}'_1)-N\kappa-\epsilon_2)} \times 2^{n(H(\underline{X}_0|\underline{Y}'_1)+\epsilon_2)} \quad (14)\\ &= 2^{n(H(\underline{X}_0,\underline{Y}'_1)-N\kappa)} \quad (15)\\ &= 2^{n(H(\underline{X}_0)-N\kappa)}, \quad (16)\end{aligned}$$

where (16) follows since $H(\underline{Y}'_1|\underline{X}_0) = 0$. Similarly, we can show that $|\mathcal{C}_{0,1}| \geq 2^{n(H(\underline{X}_0)-N\kappa-4\epsilon_2)}$.

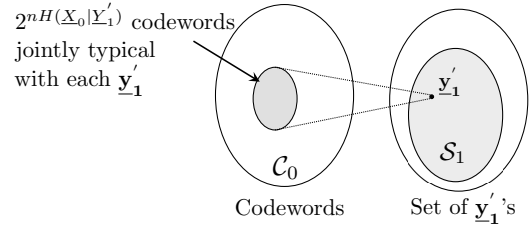

Fig. 3. Pictorial representation of pruning with respect to node 1.

If the source transmits a codeword from $\mathcal{C}_{0,1}$ in the Gaussian network, then the signal $\underline{\mathbf{y}}_1$ received by node 1 can be regarded as a noisy version of the signal $\underline{\mathbf{y}}'_1$ it would have received in the superposition network, as shown in (6). Therefore, we define a channel with input $\underline{Y}'_1$ and output $\underline{Y}_1$. Node 1 decodes by finding a vector in $\mathcal{S}_1$ that is jointly weakly $\epsilon$-typical with the received vector in the Gaussian network[1]. Since $|\mathcal{S}_1| \leq 2^{n(I(\underline{Y}'_1;\underline{Y}_1)-\epsilon_2)}$, decoding is successful with block error probability less than $\zeta$, where $\zeta \to 0$ as $n \to \infty$.

*4) Further pruning the set of codewords with respect to node 2:* There are $2^{n(H(\underline{Y}'_2|\underline{Y}'_1)\pm\epsilon_2)}$ vectors in the set of $\underline{\mathbf{y}}'_2$'s at node 2 that are jointly $\epsilon$-strongly typical with a particular $\underline{\mathbf{y}}'_1 \in \mathcal{S}_1$. Since we constructed $\mathcal{S}_2$ by randomly choosing a subset containing a $2^{-n(N\kappa+2\epsilon_2)}$ fraction of the set of all $\underline{\mathbf{y}}'_2$'s, for a large $n$, there are $2^{n(H(\underline{Y}'_2|\underline{Y}'_1)-N\kappa\pm3\epsilon_2)}$ vectors in $\mathcal{S}_2$ jointly $\epsilon$-strongly typical with each $\underline{\mathbf{y}}'_1 \in \mathcal{S}_1$. Hence, there are $2^{n(H(\underline{Y}'_1,\underline{Y}'_2)-2N\kappa\pm6\epsilon_2)}$ jointly $\epsilon$-strongly typical vectors in $\mathcal{S}_1 \times \mathcal{S}_2$ with high probability (whp) as $n \to \infty$.

Now, $2^{n(H(\underline{X}_0|\underline{Y}'_1,\underline{Y}'_2)\pm\epsilon_2)}$ codewords in $\mathcal{C}_0$ are jointly $\epsilon$-strongly typical with each $\epsilon$-strongly typical tuple in $\mathcal{S}_1 \times \mathcal{S}_2$. We iterate the procedure in the previous subsection by collecting the codewords in $\mathcal{C}_0$ which are jointly $\epsilon$-strongly typical with the $\epsilon$-strongly typical tuples in $\mathcal{S}_1 \times \mathcal{S}_2$, and denote this set by $\mathcal{C}_{0,1,2}$. Naturally, $\mathcal{C}_{0,1,2}$ is a subset of $\mathcal{C}_{0,1}$. As in (14)–(16), we obtain $|\mathcal{C}_{0,1,2}|$ is about $2^{n(H(\underline{X}_0)-2N\kappa\pm7\epsilon_2)}$ whp.

If the source transmits a codeword from $\mathcal{C}_{0,1,2}$, then nodes 1 and 2 can correctly decode to vectors in $\mathcal{S}_1$ and $\mathcal{S}_2$ respectively, with high probability for a large $n$, since $|\mathcal{S}_j| \leq 2^{n(I(\underline{Y}'_j;\underline{Y}_j)-\epsilon_2)}$ for $j \in \{1,2\}$.

---

[1]Since $\underline{\mathbf{y}}_1$ is a continuous signal, we use weak typicality to define the decoding operation. Note that strongly typical sequences are also weakly typical; hence sequences in $\mathcal{S}_1$ are weakly typical.

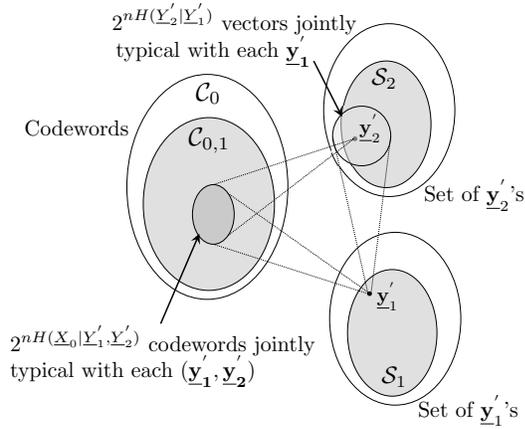

Fig. 4. Pictorial representation of further pruning with respect to node 2.

*5) Further pruning with respect to the remaining nodes:* The same procedure is repeated with respect to the remaining nodes in the network. In the end, we obtain a collection of at most $2^{n(H(\underline{X}_0) - MN\kappa + \epsilon_M)}$ and at least $2^{n(H(\underline{X}_0) - MN\kappa - \epsilon_M)}$ codewords whp, denoted by $\mathcal{C}_{0,1,\cdots,M} =: \mathcal{C}_G$, where $\epsilon_M > 0$. Note that $\epsilon_M \to 0$ as $\epsilon \to 0$. Transmission of a codeword in $\mathcal{C}_G$ results in the received signal at node $j$ in the superposition network belonging to the set $\mathcal{S}_j$.

Now, if $\mathcal{C}_G$ is used on the Gaussian network with encoding and decoding procedures at all nodes as described above, then the destination can decode to the transmitted codeword whp. Thus, on the Gaussian network, $\mathcal{C}_G$ achieves the rate

$$H(\underline{X}_0)/N - M\kappa - \epsilon_M/M = R - M\kappa - \epsilon_M/M, \quad (17)$$

where $\epsilon_M$ can be made arbitrarily small.

*6) Interleaving the codewords for general networks:* As mentioned in Sec. IV-A2, we need to slightly modify the lifting procedure for relay networks which have irregular level sets that do not permit straightforward buffering of received symbols at a node.

In this case, codewords in $\mathcal{C}_0$ are constructed by adjoining $N$ blocks of $n$ symbols each, where the first block $\underline{x}_0(1)$ consists only of the first symbols of $n$ codewords of the $(2^{NR}, N)$ code, the second block $\underline{x}_0(2)$ consists only of the second symbols of the same codewords, and so on. The source transmits $\underline{x}_0(t)$'s in the order of increasing $t$.

In the $(2^{NR}, N)$ code, let $y'_j(t)$, $t = 1, \ldots, N$, denote the $t$-th symbol received by node $j$. We adjoin the $t$-th received symbols from $n$ uses of the code to construct $\underline{y}'_j(t)$. Since $x_j(t)$, the $t$-th symbol transmitted by node $j$, is a function of $\{y'_j(p)\}_{p=1}^{t-1}$, node $j$ can construct $\underline{x}_j(t)$, vector consisting of the $t$-th transmit symbols from $n$ uses of the code, after receiving $\{\underline{y}'_j(p)\}_{p=1}^{t-1}$, .

Essentially, we interleave the symbols from $n$ uses of the same code to ensure that the nodes can buffer their receptions.

In order to lift the coding scheme to the Gaussian network, we prune $\mathcal{C}_0$ by randomly picking a $2^{-n(\kappa+2\eta)}$-fraction of the set of $\epsilon$-strongly typical $\underline{y}'_j(t)$, for all $t$, for all $j$, and collecting the codewords jointly $\epsilon$-strongly typical with them to form $\mathcal{C}_G$.

In the Gaussian network, each node buffers its reception for $n$ time units, decodes to the appropriate $\underline{y}'_j(t)$, constructs $\underline{x}_j(t+1)$, transmits it on the next $n$ time units. The destination decodes individual $n$-length blocks to get $\underline{y}'_M(t)$, $t = 1, 2, \ldots, N$, and decodes to a codeword in $\mathcal{C}_G$ after de-interleaving $\{\underline{y}'_M(t)\}$.

## V. MIMO NETWORKS

The results in Theorem 3.1 can be extended to MIMO networks, where nodes have multiple transmit or receive antennas. In that case, $\kappa$ determining the bounded gap is a function of the number of transmit and receive antennas at the various nodes as well as the number of nodes in the network.

For simplicity, consider a network where every node has two transmit and two receive antennas. All transmitted and received signals in both the Gaussian and discrete superposition model for this network are a pair of vectors. For example, node $j$'s received signal in the Gaussian model is $\underline{\mathbf{y}}_\mathbf{j} = [\underline{\mathbf{y}}_{\mathbf{j},\mathbf{1}}, \underline{\mathbf{y}}_{\mathbf{j},\mathbf{2}}]$. The channel on a particular wireless link $(i, j)$ is specified by four channel gains, $\{h_{ij}^{k,l}\}$, where $k \in \{1, 2\}$ indexes the transmit antennas of $i$ and $l \in \{1, 2\}$ indexes the receive antennas of $j$. Assuming that the noises at both receive antennas are distributed as $\mathcal{CN}(0, 1)$, we have

$$\underline{\mathbf{y}}_{\mathbf{j},\mathbf{l}} = h_{ij}^{1,l}\underline{\mathbf{x}}_{\mathbf{i},\mathbf{1}} + h_{ij}^{2,l}\underline{\mathbf{x}}_{\mathbf{i},\mathbf{2}} + z_{j,l}, \, l = 1, 2. \quad (18)$$

The arguments in Sections III and IV can be applied unchanged to this model by replacing every vector with its corresponding tuple of vectors. However, $\kappa$ is different:

$$\begin{aligned}\kappa &:= 2\log(24M - 2) + 22 \\ &\geq 2 \max_{j \in \mathcal{V} \setminus \{0\}} H([V_{j,1}]) + H([Z_{j,1}]) + H(C_{j,1}),\end{aligned}$$

where $v_{j,1}$ and $c_{j,1}$ are defined appropriately with respect to $\underline{\mathbf{y}}_{\mathbf{j},\mathbf{1}}$ (see (6)). The above bound exploits the choice of the same number of antennas at every node. Otherwise, we can use the maximum number of antennas at any node in the network to arrive at a worst-case upper bound.

## VI. CONCLUDING REMARKS

We have demonstrated a procedure to lift *any* coding scheme from the discrete superposition model to the Gaussian network. The next step that is of interest is to develop good *structured* codes for the superposition model so that we can then lift them to obtain good structured codes for the Gaussian network.

## ACKNOWLEDGEMENT

The authors wish to acknowledge helpful discussions with Hemant Kowshik on the proof of Theorem 3.1.